\documentclass[graybox, envcountchap,usenatbib]{svmult}

\usepackage{mathptmx}        
\usepackage{amsmath}
\usepackage{amssymb}
\usepackage{color}
\usepackage{helvet}          
\usepackage{courier}         
\usepackage{dirtree}

\usepackage{makeidx}        
\usepackage{graphicx}        
\usepackage{subfig}

\usepackage{multicol}        
\usepackage[bottom]{footmisc}

\usepackage{hyperref}        
\hypersetup{colorlinks=true,urlcolor=blue,citecolor=blue}

\usepackage[misc]{ifsym}

\makeindex             

\usepackage[english]{babel}
\usepackage[square,numbers]{natbib}

\begin{document}


\title{The pursuit of the Hubble Constant using Type II Supernovae}
\author{Thomas de Jaeger \& Llu\'is Galbany}
\institute{Thomas de Jaeger (\Letter) \at LPNHE, CNRS/IN2P3 \& Sorbonne Universit\'e, 4 place Jussieu, 75005 Paris, France\\ \email{thomas.dejaeger@lpnhe.in2p3.fr}
\and Llu\'is Galbany (\Letter) \at Institute of Space Sciences (ICE-CSIC), Campus UAB, Can Magrans S/N, 08193 Barcelona, Spain \\
Institut d’Estudis Espacials de Catalunya (IEEC), 08034 Barcelona, Spain.\\
\email{lg@csic.es}}
%
%
\maketitle

\abstract{
The use of multiple independent methods with their own systematic uncertainties is crucial for resolving
the ongoing tension between local and distant measurements of the Hubble constant ($H_{0}$).
While type Ia supernovae (SNe~Ia) have historically been the most widely used distance indicators, recent studies have shown that type II supernovae (SNe~II) can provide independent measurements of extragalactic distances with different systematic uncertainties. 
Unlike SNe~Ia, the progenitors of SNe~II are well understood, arising from the explosion of red supergiants in late-type galaxies via core-collapse. 
While SNe~II do not exhibit the same level of uniformity in peak luminosity as SNe~Ia, their differences can be calibrated using theoretical or empirical methods.
Overall, this chapter presents a comprehensive overview of the use of SNe~II as extragalactic distance indicators, with a particular focus on their application to measuring $H_0$ and addressing the Hubble tension. 
We describe the underlying theory of each method, discuss the challenges associated with them, including uncertainties in the calibration of the supernova absolute magnitude, and present a comprehensive list of the most updated Hubble constant measurements. 
}


\section{Introduction}

Thermonuclear explosions of a carbon-oxygen white dwarf in multi-star systems, known as Type Ia supernovae (SNe Ia), are among the best distance indicators \citep{Phillips:1993ng,Hamuy:1996ss,Riess:1996pa}. For more than three decades, they have been used to measure extragalactic distances with an accuracy of 5-10\% \citep{Betoule:2014iwm,Brout:2022vxf} and have been fundamental to identify a tension between local \citep{Riess:2021jrx} and distant \citep{Planck:2018vyg} measurements of the Hubble constant ($H_0$). There are two complementary ways to solve this tension: first, increasing the level of precision of the different methods, and second, developing as many independent methods as possible with their own systematic uncertainties. In this chapter, as an independent approach, we describe the use of Type II supernovae (SNe II) to estimate $H_0$. 

After being overshadowed by the better standardized SNe~Ia, recent studies have demonstrated that SNe~II can be used to measure extragalactic distances and have a role to play in the ``$H_0$ tension.'' Observationally, their optical spectra are characterized by the presence of strong Balmer lines exhibiting broad P-Cygni profiles (e.g., \citep{Filippenko:1997ub,Gal-Yam:2016yms}), and their light curves display a plateau of varying steepness and length (\citep{Barbon79,Anderson:2014hta,Galbany:2015pbi,deJaeger:2019nya}) owing to hydrogen recombination. Among all the supernova types, SN~II is the one for which the physics and the nature of their progenitors are the most well-understood. Unlike SNe~Ia, for which no clear progenitors have been yet identified, SNe~II have dozens of direct progenitor detections \citep{Smartt:2015sfa}. It is now accepted that their progenitors have retained a significant fraction of their Hydrogen envelopes and arise from the explosion of red supergiants in late-type galaxies via core collapse. The understanding of the explosion mechanisms of SN II itself has also made remarkable progress in the past few decades \citep{Woosley:1995ip,Janka:2012wk}.

Although SNe~II are not perfect standard candles and display a large range of peak luminosities (\citep{Anderson:2014hta}), they have the potential to be excellent distance indicators with different systematics than SNe~Ia. Their extrinsic differences can be calibrated using theoretical or empirical distance measurement methods such as the expanding photosphere method (EPM; \cite{Kirshner:1974ghm}), the spectral expanding atmosphere method (SEAM; \cite{Baron:2004wb,Dessart:2007rt}), the tailored EPM (\cite{Vogl:2020thesis}), the standard-candle method (SCM; \cite{Hamuy:2002tj}), the photospheric-magnitude method (PMM; \cite{Rodriguez:2014,Rodriguez:2019}), and the only method based solely on photometry, the photometric-color method (PCM; \cite{dejaeger:2015dqj,deJaeger:2016cev,DES:2020wgl}). 

Among this wide variety of methods to measure distances from SN~II, here, we review only those used to measure $H_0$. For each method, we will describe the underlying theory behind them and present the most updated Hubble diagram together with a nearly exhaustive list of their $H_0$ measurements.

\section{EPM: Expanding Photosphere Method}

The Expanding Photosphere Method is a geometric-based method based on the approach used by \cite{Baade:1926cj} to measure distances for pulsating stars. The distance ($D$) is directly obtained by measuring the physical photosphere radius ($R_{phot}$) and the angular size ($\theta$) of the SN~II. Unfortunately, extragalactic SNe are unresolved and those parameters cannot be measured directly, they need to be derived by making some assumptions. First, supported by the fact that SNe II show a low polarization degree during the plateau phase (\cite{Leonard:2001vi}), we assume a spherically symmetric photosphere. Second, the ejecta is freely expanding, so we can determine the photosphere radius from its velocity ($v$) and time since explosion ($t-t_{0}$): $R_{phot}(t) = v(t-t_{0}) + R_{0}$. Due to a fast expansion, it is safe to assume that the initial radius ($R_{0}$ at $t=t_{0}$ ) is negligible after a few days. The velocities of the photosphere are determined through the minimum flux of the absorption component of the P-Cygni line profile of weak lines such as those of Fe II ($\lambda$5018,5169). Finally, to derive the theoretical angular size $\theta$, we assume that the photosphere radiates as a blackbody (BB), i.e., the thermalization layer from which the thermal photons originated (R$_{therm}$) and the photosphere have the same radius. Therefore, we have $4\pi D^{2} f_{\nu} = 4 \pi R_{therm}^{2} \pi B_{\nu}(T_{c}) 10^{-0.4A_{\lambda}}$, where B$_{\nu}(T_{c})$ is the Planck function at color temperature $T_{c}$ of the BB radiation estimated by fitting Planck curves to observed broadband magnitudes, A$_{\lambda}$ the total dust extinction (host galaxy and Milky Way), and $f_{\nu}$ is the flux received at Earth. Finally, for $z<<1$, $\theta$ can be written as,
\begin{eqnarray}
\theta=\frac{R_{phot}}{D} \overset{\textbf{BB}}{=} \frac{R_{therm}}{D}= \sqrt{\frac{f_{\nu}}{\pi B_{\nu}(T_{c}) 10^{-0.4A_{\lambda}}}}.
\label{eq:EPM}
\end{eqnarray}
Combining $R(t) = v(t-t_{0})$ and the above equation implies that $t = D(\theta/v) + t_{0}$, i.e., the distance $D$ corresponds to the slope of a straight line made in the $t$ versus ($\theta/v$) plot and the explosion date to the y-intercept. 

The EPM was first used by \cite{Kirshner:1974ghm} to
calculate distances of two SNe~II (69L and 70g). They also obtained the first $H_0$ from SN~II and found a value of $H_0= 65 \pm 15 $\,km\,s$^{-1}$ Mpc$^{-1}$. However, a few years later, \cite{Wagoner:1981} noticed that the photosphere radiation deviates from that of a BB. The reason is that the thermalization layer from which the BB radiation is generated is deeper than the photosphere. The result is that the photosphere radiates less strongly than a BB of the same color temperature and therefore, appears diluted. To take into account this effect, the EPM was refined by incorporating a dilution factor ($\xi$) in equation \ref{eq:EPM} which parameterized the relative position of the two surfaces ($R_{phot}= R_{therm}/\xi$). In principle, the dilution factor should account for all deviations of the spectrum from BB emission, including the chemical composition or density structure of the progenitor star. However, theoretical dilution factors computed from non-LTE models with a wide range of properties have shown that they depend only on the color temperature (\cite{Eastman:1996aaa,Dessart:2005gg}).

\begin{figure*}[!t]
	\includegraphics[width=\textwidth]{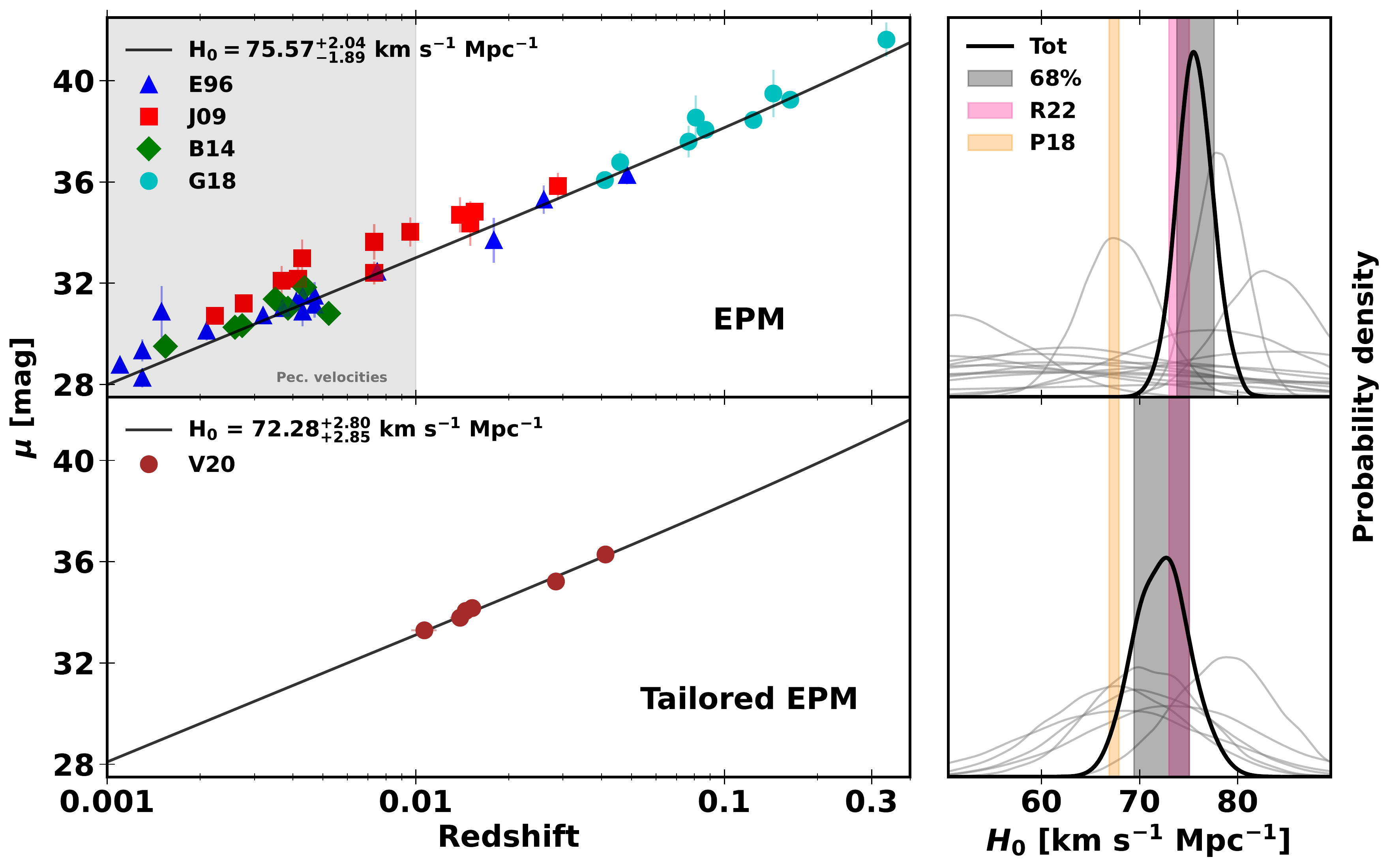}
 \caption{\textit{Left Top:} Figure updated from figure of \cite{Gall:2017gva} representing the Type II supernova Hubble diagram using the EPM. The EPM was applied to the data taken from \cite{Eastman:1996aaa} (blue triangles; E96), \cite{Jones:2009vu} (red squares; E96), \cite{Bose:2014sza} (green diamonds; B14), and \cite{Gall:2017gva} (cyan dots; G18). \textit{Left Bottom:} Type II supernova Hubble diagram of six objects using the distance luminosities derived by \cite{Vogl:2020thesis} (brown dots; V20) when applying the tailored EPM. \textit{Right Top} and \textit{Right bottom} show the probability distributions of $H_{0}$ for the individual SNe in grey and the combined estimate in black. We include the value from local measurement \citep{Riess:2021jrx} (pink filled region) and from the high-redshift results \cite{Planck:2018vyg} (orange filled region) for comparison.}
\label{fig:HD_EPM_tot}
\end{figure*}

\cite{Schmidt:1994fu} were the first to apply this refined method to a large sample of SN~II and derive a $H_0$. Using 16 objects and preliminary values of the dilution factors computed by \cite{Eastman:1996aaa}, they obtained a value of 73 $\pm ~6~{\rm (stat)} \pm 7~{\rm (sys)}$\,km\,s$^{-1}$ Mpc$^{-1}$. Then, using 9 well-observed SNe~II, with a slightly modified version of \cite{Eastman:1996aaa} dilution factors, and different filter combinations, \cite{Hamuy:2001yt} obtained an average value $H_{0}=$71 $\pm ~9~{\rm (stat)}$\,km\,s$^{-1}$ Mpc$^{-1}$. Based on \cite{Hamuy:2001yt} work, \cite{Leonard:2003an} derived a value of $H_{0}=$57 $\pm ~7~{\rm (stat)}\pm$ ~13~{\rm (sys)} \,km\,s$^{-1}$ Mpc$^{-1}$. Two other works have used the EPM to measure the Universe expansion rate. \cite{Jones:2009vu} constructed a Hubble diagram with 12 well-observed SNe~II using dilutions factors from \cite{Eastman:1996aaa} and \cite{Dessart:2005gg}. They obtained a value of $H_{0}=$100.5 $\pm ~8.4~{\rm (stat)}$\,km\,s$^{-1}$ Mpc$^{-1}$ and $H_{0}=$56.9 $\pm ~4.2~{\rm (stat)}$\,km\,s$^{-1}$ Mpc$^{-1}$, respectively. The dilution factor choice generates a large systematic difference in the $H_{0}$ values. The factors calculated by \cite{Dessart:2005gg} are systematically larger than \cite{Eastman:1996aaa} which implies that the EPM distances are also $\sim$ 10--20\% larger. The difference between the two sets of dilution factors is unclear but it is an important source of concern for using the EPM to measure distances. \cite{Gall:2016qvq} also found different values for $H_{0}$ depending on the dilution factors. For one object (13eq) in the Hubble flow ($z=0.041$), they obtained a value of $H_{0}=$ 83 $\pm ~10~{\rm (stat)}$\,km\,s$^{-1}$ Mpc$^{-1}$ and $H_{0}=$ 76 $\pm ~9~{\rm (stat)}$\,km\,s$^{-1}$ Mpc$^{-1}$ using dilution factors from \cite{Hamuy:2001yt} and \cite{Dessart:2005gg} respectively. Finally, \cite{Gall:2017gva} have successfully extended the EPM to high-z and measured the distances to 10 SNe~II with a $z>0.04$. 
Using these distance measurements\footnote{We have not considered LSQ13cuw, which is an outlier in the EPM and SCM Hubble diagrams (see Figure 7 from \cite{Gall:2017gva})}, 
together with those of seven other SNe~II in the Hubble flow ($z>0.01$) from the literature, we construct a combined probability distribution from the individual $H_{0}$ posteriors of each SN~II using Gaussian kernel-density estimates (see Figure \ref{fig:HD_EPM_tot})\footnote{using $H_{0}= cz/Dl [1+0.5(1-q_{0})z]$ with q$_{0}$=-0.55}. We obtained $H_{0}=$ 75.53$^{+2.07}_{-1.97}~{\rm (stat)}$ and $H_{0}=$ 84.70$^{+2.28}_{-2.21}~{\rm (stat)}$ using dilution factors from \cite{Hamuy:2001yt} and \cite{Dessart:2005gg}, respectively. \\

\noindent
\textbf{Pros}: Absolute distances, i.e., do not require any external calibration. Less sensitive to uncertainties in the dust extinction.\\
\textbf{Cons}: Dilution factors: different sets of dilution factors yield differences of $\sim$ 20\% in the inferred distance.\\
\textbf{Requirements}: Optical photometry (at least $BVI$), at least two optical spectra, and dilution factors.\\
\textbf{Future}: New dilution factors based
on independent numerical methods (e.g. \cite{Vogl:2018ckb}).

\section{Tailored EPM}

To avoid systematic biases introduced by the use of the dilution factors with the EPM, the Spectral
Expanding Atmosphere Method (SEAM; \cite{Baron:96,Baron:2000kq,Baron:2004wb} and the tailored EPM (\cite{Dessart:2005gg,Vogl:2020thesis}) have been developed. Both methods take advantage of the full information contained in the observed spectra through detailed spectral fitting to infer the chemical composition, the density profile, the temperature, the ejecta velocity, and other parameters of the SN. Afterwards, because the spectral energy distribution is known, one can determine the absolute magnitude in any bands from the calculated synthetic spectra. Then a comparison with to the observed apparent magnitudes allows us to obtain the distance modulus.

Mainly three radiative-transfer codes have been used to fit observed spectra: \textit{PHOENIX} (\cite{Hauschildt:1998jw,Baron:2004wb}), \textit{CMFGEN} (\cite{Dessart:2005gg}, and a modified version of \textit{Tardis} (\cite{Kerzendorf:2014,Vogl:2018ckb}). All the codes take into account the departure from the local thermodynamic equilibrium (LTE) seen in low-density, scattering-dominated SN II atmospheres via the use of a non-LTE approximation for the ionization and excitation treatment of the species. However, radiative-transfer simulations are complicated and time-consuming, ranging from minutes to days on large supercomputers depending on the number of free parameters. Additionally, it requires high signal to noise spectrum sequence but only recently such data have been made publicly accessible. For these reasons, only three well-observed SNe~II in the local Universe (99em, 05cs, 06bp) have been studied using detailed radiative-transfer models (\cite{Baron:2004wb,Dessart:2007rt}). The authors provided excellent fits between the observed and synthetic spectra, with a level of accuracy on the distances $\sim$ 10\%. However, none of those SNe~II are enough distant to derive a reliable $H_{0}$.

Recently, radiative-transfer simulations have regained popularity thanks to the development of a fitting method using a spectral emulator (\cite{Vogl:2019fhc}) to predict the SN II spectra and magnitudes. The emulator has been trained on a large grid of spectra calculated with \textit{Tardis}, and can produce synthetic spectra in only 10$^{-2}$s instead of 10$^{5}$s. It has been built in two steps: (1) data dimensionality reduction by Principal Component Analysis (PCA), (2) Gaussian processes to interpolate the spectra within the PCA space and to predict preprocessed, dimensionality-reduced spectra for new sets of input parameters including: photospheric temperature and velocity, metallicity, time since explosion, and the exponent of the density profile. Finally, a synthetic spectrum is obtained by reversing the preprocessing procedure and then compare to the observations. \cite{Vogl:2019fhc} demonstrated the emulator predictions match the \textit{Tardis} simulations with a precision of better than 99\%.  \cite{Vogl:2020thesis} applied their new method to six low-redshift SNe~II (03bn, 06it, 10id, 12ck, 13fs, and iPTF13bjx) and constructed the Hubble diagram displayed in Figure \ref{fig:HD_EPM_tot} (bottom left). From those six distances, they were able to derive a value of $H_{0}=$ 75.28$^{+2.80}_{-2.85}~{\rm (stat)}$. This result is currently limited by a small sample size and 
low redshift range ($z<0.04)$,
and it is affected by the systematic uncertainties of atmosphere models. However, this is a promising method which is independent of local calibrations. Also, it has recently been checked using sibling SNe~II, i.e., SNe~II in the same host galaxy. With four different host galaxies (8 SNe~II), \cite{Csornyei:2023rpw} found consistent distances within the errors for each of the SN sibling pairs and with a precision better than 5\% for two hosts. Finally, the adH0cc collaboration (accurate determination of $H_{0}$ with core-collapse supernovae\footnote{https://adh0cc.github.io/}) has now gathered observations for 26 SNe~II in the Hubble flow ($z >$ 0.03). This new dataset will provide the basis for a highly-competitive $H_{0}$ measurement.\\

\noindent
\textbf{Pros}: Absolute distances, i.e., do not require any external calibration. Unlike the EPM, dilution factors are not needed.\\
\textbf{Cons}: Small sample size (only 6 objects) and 
low redshift range ($z<0.04$).
Assume fractional uncertainties for $\theta/v_{ph}$.\\
\textbf{Requirements}: A well-defined explosion date, optical photometry (at least $BVI$) interpolated to the spectral epochs, multiple well-calibrated spectra in the first month after the explosion, and spectral emulator to predict synthetic spectra.\\
\textbf{Future}: Increase the sample of SN~II in the Hubble flow. Better treatment of the uncertainties.

\begin{figure*}[!t]
	\includegraphics[width=\textwidth]{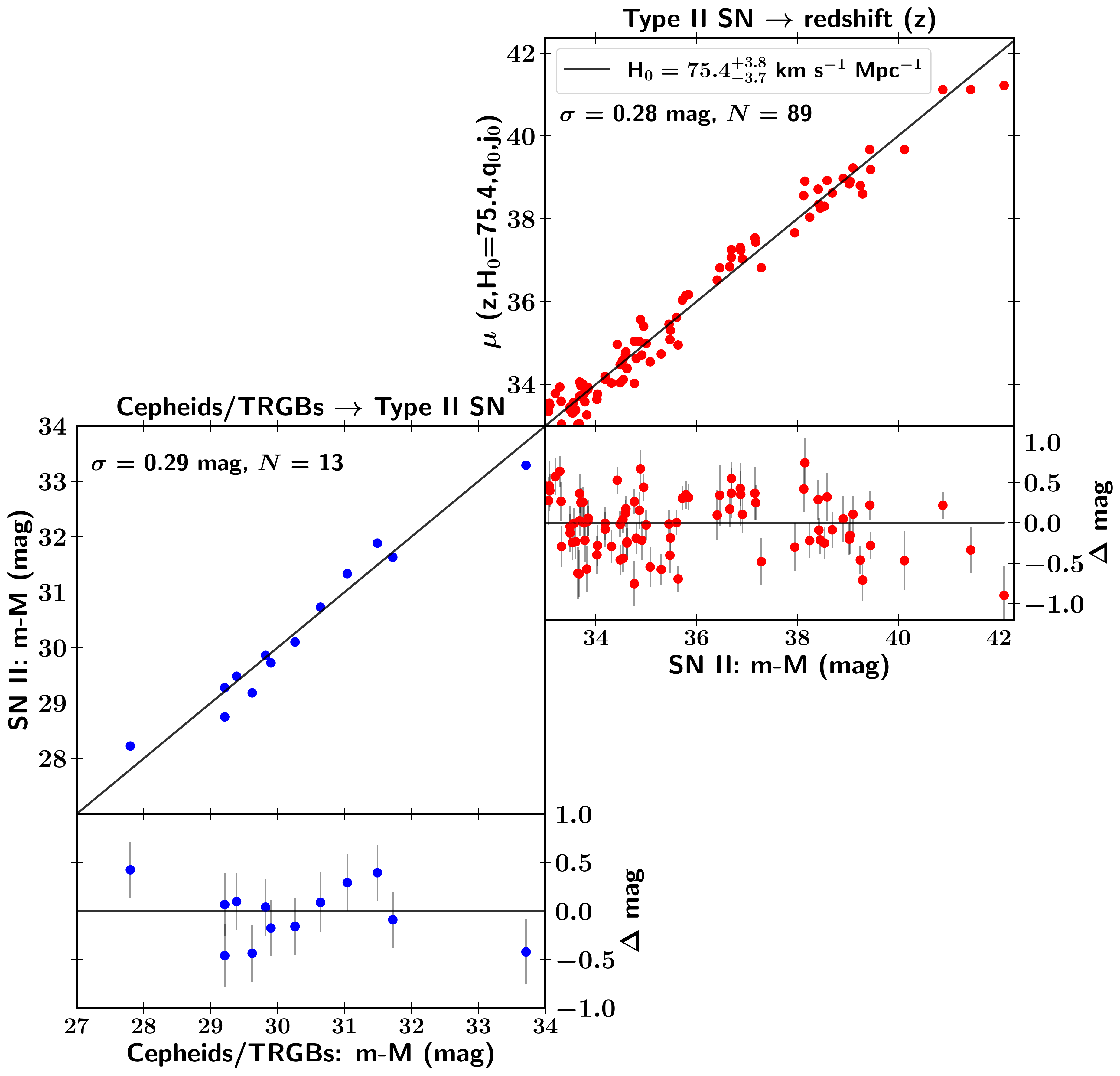}
 \caption{Figure from \cite{deJaeger:2022lit} representing the last two rungs of the distance ladder with the SCM. For each rung, distances on the abscissa serve to calibrate a relative distance on the ordinate. Blue dots represent geometric, Cepheid, or TRGB distances used to calibrate SN~II through the determination of the $i$-band absolute magnitudes (\textit{Bottom left}). Red dots are the SNe II in the Hubble flow used to constrain $H_0$ (\textit{Top right}).}
\label{fig:HD_SCM}
\end{figure*}

\section{SCM: Standard Candle Method}\label{sec:SCM}

Unlike the EPM and its tailored version, the Standard Candle Method (SCM) is an empirical method. First developed by \cite{Hamuy:2001thesis}, it has been later theoretically explained by \cite{Kasen:2009ai}. SCM is based on the correlation between the SN absolute magnitude and the photospheric expansion velocity determined from the blueshift of the Fe II lines ($\lambda$5018). Such a correlation is expected when the SN is on the plateau phase, i.e., between 30--80 days after the explosion (\cite{Anderson:2014hta}). This phase is physically well-understood and is only due to a change in opacity and density in the outermost layers of the SN. A few weeks after the explosion, the ionized hydrogen present in the outermost layers of the progenitor starts to recombine when the temperature drops and levels of the hydrogen recombination temperature ($\sim$5,000 K). During this phase, the light curve is powered by the sudden release of energy caused by the recession of the photosphere in mass that corresponds to a fixed radius in space. Therefore, as the temperature and the radius are nearly constant, the luminosity also appears constant. Consequently, in more luminous SNe~II, the hydrogen recombination front will be fixed at a larger radius, and therefore, assuming a homologous expansion, the photosphere will be maintained at higher velocities.

\cite{Hamuy:2003tc} applied this technique to 24 SNe~II in the Hubble flow and reduced the Hubble diagram scatter from 0.83 mag to only 0.32 mag, i.e., 15\% in distances. Distances derived using the SCM are not absolute and must be calibrated using primary distance indicators like Cepheids or Tip Red Giant Branch (TRGB). Using four SNe~II (68L, 70G, 73R, 99em) with precise Cepheid distances, \cite{Hamuy:2003tc} derived a $H_0$ of 75 $\pm$ 7 km s$^{-1}$ Mpc$^{-1}$. With the same technique but with only one SN (99em) calibrated via Cepheid, \cite{Leonard:2003an} obtained a smaller $H_0$ of 59 $\pm ~3~{\rm (stat)} \pm 11~{\rm (sys)}$.

Recently, many others studies have refined the SCM to simply its use and to extend it to higher redshifts. First, \cite{Nugent:2006mwe} added an extinction correction (intrinsically brighter SNe II are bluer) like for SN~Ia cosmology ($\beta$ $\times$ color) rather than relying on a method to measure the host-galaxy extinction. Second, he adjusted the velocity of the Fe II lines with a power-law decline, reducing the need of obtaining several spectra around the middle of the plateau phase. Third, \cite{Nugent:2006mwe,Poznanski:2008yg,Takats:2011ti,deJaeger:2016cev} have proposed to use the H$\beta$ $\lambda$4861 absorption line which has the advantage of being stronger than Fe II lines ($\lambda$5018), and therefore easier to measure. Finally, because the measurement of the velocity can be difficult for noisy even for the strong H$\beta$ $\lambda$4861 absorption line, \cite{Poznanski:2008yg} and \cite{deJaeger:2017qxx} have developed a method based on the cross-correlation between the observed spectra and a library of high S/N SN II spectra. Velocities from direct measurement or cross-correlation techniques
have shown a good agreement with a dispersion of only 400 km s$^-1$ (\cite{deJaeger:2017qxx}). Therefore, finally, using the SCM, the corrected magnitude is written as,
\begin{eqnarray}
m_{\rm corr}=\mathrm{m}+\alpha\, \mathrm{\log_{10}}(v_\mathrm{H\beta}) - \beta color,
\label{m_model}
\end{eqnarray}
where $m$ is the apparent magnitude in a given passband around 50\,d after the explosion, $v_\mathrm{H\beta}$ is the velocity measured using H$\beta$ absorption from an optical spectrum, $\alpha$ and $\beta$ are nuisance parameters. The fine-tuned version of the SCM have been used by several studies (\cite{Poznanski:2008yg,Poznanski:2010my,DAndrea:2009lzz,OlivaresE:2010idm,dejaeger:2015dqj,deJaeger:2016cev,Gall:2017gva,deJaeger:2017qxx,DES:2020wgl}) to measure SN~II distances in the Hubble flow with a precision of 12--15\% in distance. However, only three works have applied this updated version to measure $H_0$. First, \cite{OlivaresE:2010idm} used two SNe~II with Cepheid distances (99em, 04dj) and obtained a value of 71 $\pm$ 12 km s$^{-1}$ Mpc$^{-1}$, 69 $\pm$ 16 km s$^{-1}$ Mpc$^{-1}$, and 70 $\pm$ 16 km s$^{-1}$ Mpc$^{-1}$ depending on the passband used ($B$, $V$, and $I$, respectively). More recently, using seven objects (99em, 99gi, 05ay, 05cs, 09ib, 12aw, and 13ej) with
Cepheid or TRGB independent host-galaxy distance measurements, \cite{deJaeger:2020zpb} found a $H_0$ value of $75.8^{+5.2}_{-4.9}$ (stat) $\pm$ 2.8 (sys) \,km\,s$^{-1}$\,Mpc$^{-1}$, which differs by $1.4\sigma$ from the high-redshift result (\cite{Planck:2018vyg}). Thanks to six additional calibrators (04et, 08bk, 14bc, 17eaw, 18aoq, 20yyz) and a better analysis of the systematic errors, \cite{deJaeger:2022lit} were able to obtain a $H_0$ with an uncertainty of 5\%,  $75.4^{+3.8~{\rm(stat)}}_{-3.7~{\rm (stat)}} \pm 1.5~{\rm (sys)}$\,km\,s$^{-1}$\,Mpc$^{-1}$. In Figure \ref{fig:HD_SCM}, the last two rungs of the distance ladder from their latest work are displayed. Their value is consistent with the local measurement \citep{Riess:2021jrx} but differs by $2.0\sigma$ from the high-redshift results \cite{Planck:2018vyg}. Additionally, using seven Cepheids or five TRGB, they derived values which differ by $<1.0\sigma$ (i.e., 4.5\,kms$^{-1}$\,Mpc$^{-1}$), suggesting that neither Cepheids nor TRGB are the source of the ``$H_0$ tension.''. It is important to note that one SN~II (14bc) was discovered in NGC~4258 for which we have geometric distance. Therefore, we could use 14bc to directly calibrate the SN~II in the Hubble flow, without the need for any calibrators, i.e., it will be just a two-rungs distance ladder. However, the uncertainties on its SCM distance and mostly the uncertainty on the ejecta expansion velocity are too large to constrain $H_0$.\\

\noindent
\textbf{Pros}: Straightforward simple method and has been applied at high redshift.\\
\textbf{Cons}: Needs to be calibrated. Common reddening law for all SNe.\\
\textbf{Requirements}: Optical photometry (at least two bands) up to 50 days after the explosion, one optical spectrum during the plateau phase, a well-defined explosion date ($\sigma_{T_{exp}}<$10 days), calibrators (Cepheids, TRGB) in the same host.\\
\textbf{Future}: Increase the number of calibrators, and find new correlations to decrease the scatter in the Hubble diagram (e.g host galaxy environment).

\section{PMM: Photospheric Magnitude Method}

\begin{figure*}[!t]
	\includegraphics[width=\textwidth]{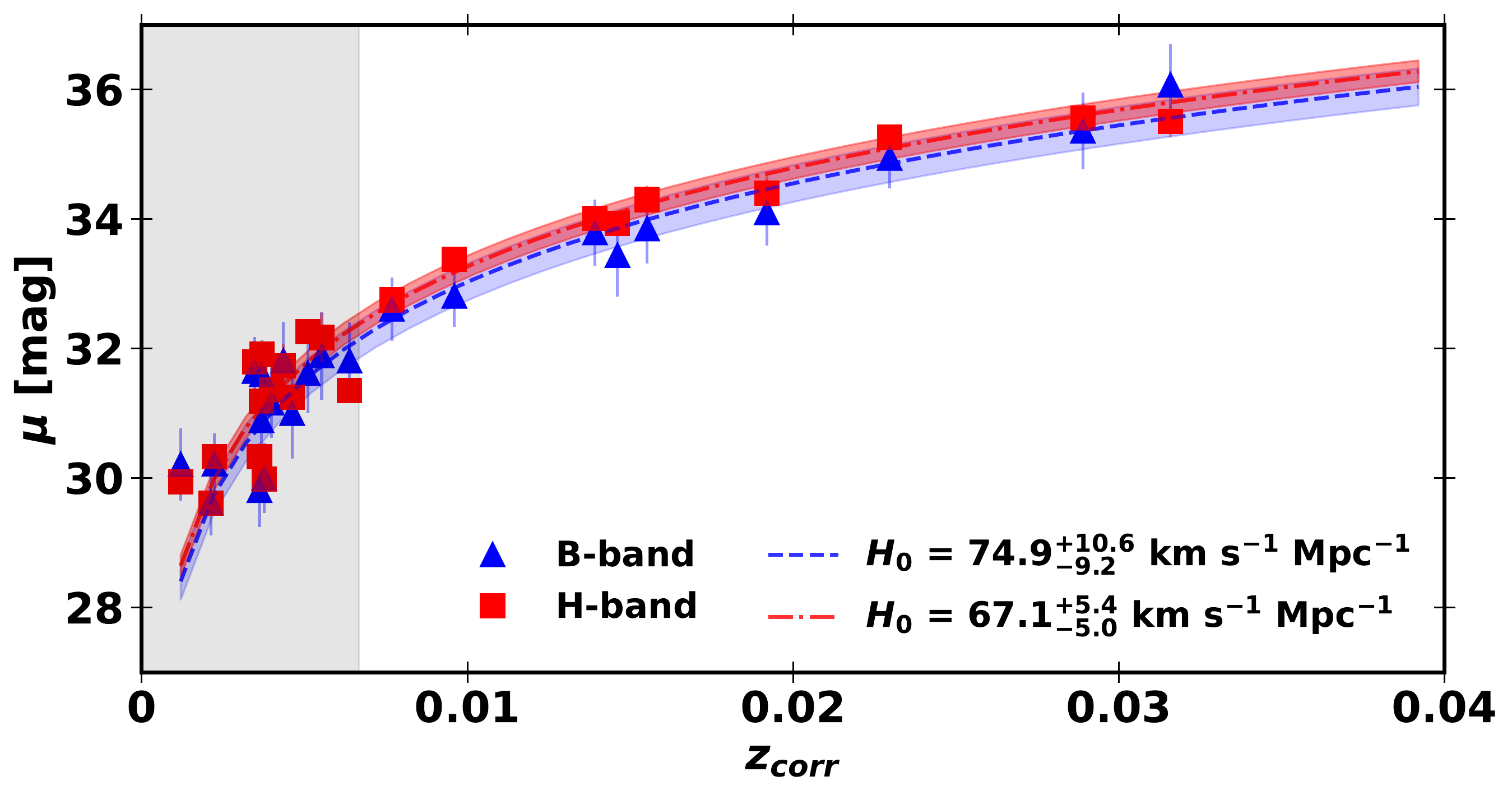}
 \caption{Figure updated from figure of \cite{Rodriguez:2019} representing SNe~II distance moduli derived with the PMM for the $B$ (blue triangles) and $H$ (red squares) bandpass versus the redshift corrected for peculiar velocities. Red and blue lines correspond to the Hubble law fits to only the SNe~II at $cz$ $>$ 2000 km s$^{-1}$ (shaded region).}
\label{fig:HD_PMM}
\end{figure*}

\begin{table}
\small
\caption{$H_0$ values from Type II Supernovae}
\centering
\label{tab:1}       
\begin{tabular}{p{4.1cm}p{3.5cm}p{2cm}p{1.5cm}}
\hline\noalign{\smallskip}
$H_0$ & Method$+$ & year & Reference  \\
km s$^{-1}$ Mpc$^{-1}$ & dilution factors or N$_{cal}$ &  &   \\
\noalign{\smallskip}\svhline\noalign{\smallskip}
60 $\pm ~15~{\rm (stat)}$ & EPM & 1974 &  \cite{Kirshner:1974ghm}\\
73 $\pm ~6~{\rm (stat)} \pm 7~{\rm (sys)}$ & EPM$+$\cite{Eastman:1993aaa} & 1994 &  \cite{Schmidt:1994fu}\\
71 $\pm ~9~{\rm (stat)}$ & EPM$+$\cite{Hamuy:2001yt} & 2001 &  \cite{Hamuy:2001yt}\\
57 $\pm ~7~{\rm (stat)} \pm 13~{\rm (sys)}$ & EPM$+$\cite{Hamuy:2001yt} & 2003 & \cite{Leonard:2003an}\\
100.5$\pm ~8.4~{\rm (stat)}$ & EPM$+$\cite{Eastman:1996aaa} & 2009 &  \cite{Jones:2009vu}\\
59.6$\pm ~4.2~{\rm (stat)}$ & EPM$+$\cite{Dessart:2005gg} & 2009 &  \cite{Jones:2009vu}\\
83 $\pm ~10~{\rm (stat)}$ & EPM$+$\cite{Hamuy:2001yt} & 2016 &  \cite{Gall:2016qvq}\\
76 $\pm ~9~{\rm (stat)}$ & EPM$+$\cite{Dessart:2005gg} & 2016 &  \cite{Gall:2016qvq}\\
84.70$^{+2.28}_{-2.21}~{\rm (stat)}~\pm 10.47~{\rm (sys)}$ & EPM$+$\cite{Hamuy:2001yt} & 2023 &  This work\\
75.57$^{+2.04}_{-1.89}~{\rm (stat)}~\pm 13.46~{\rm (sys)}$ & EPM$+$\cite{Dessart:2005gg} & 2023 &  This work\\
72.28$^{+2.80}_{-2.85}~{\rm (stat)}~\pm 3.79~{\rm (sys)}$ & tailored EPM & 2020 &  \cite{Vogl:2020thesis}\\
\noalign{\smallskip}\svhline\noalign{\smallskip}
74.9$^{+10.6}_{-9.2} \rm(stat)$ & PMM$+$4$+$$B$ & 2019 & \cite{Rodriguez:2019}\\
67.1$^{+5.4}_{-5.0} \rm(stat)$ & PMM$+$4$+$$H$ & 2019 & \cite{Rodriguez:2019}\\
\noalign{\smallskip}\svhline\noalign{\smallskip}
59 $\pm ~3~{\rm (stat)} \pm 11~{\rm (sys)}$ & SCM$+$1 & 2003 & \cite{Leonard:2003an}\\
75 $\pm ~7~{\rm (stat)} $ & SCM $+$4& 2003 & \cite{Hamuy:2003tc}\\
71 $\pm ~12~{\rm (stat)} $ & SCM$+$2 & 2010 & \cite{OlivaresE:2010idm}\\
75.8$^{+5.2}_{-4.9} \rm(stat) \pm 2.8~{\rm (sys)}$ & SCM$+$7 & 2020 & \cite{deJaeger:2020zpb}\\
75.4$^{+3.8}_{-3.7} \rm(stat) \pm 1.5~{\rm (sys)}$ & SCM$+$13 & 2022 & \cite{deJaeger:2022lit}\\
\noalign{\smallskip}\hline\noalign{\smallskip}
\end{tabular}
\end{table}

\cite{Rodriguez:2014} developed another method to measure SN~II distances and build Hubble diagram with a scatter of 0.15 mag. This method is a generalization of the SCM (Section \ref{sec:SCM}) at any epoch during the photospheric phase. For a given bandpass $X$, the absolute magnitude ($M_{X}$) at an epoch $t_{i}$ (since the explosion $t_{0}$) can be written as,
\begin{eqnarray}
\mathrm{M_{X,t_{i}}}=a_{X,t_{i}}- 5log_{10}(v_{ph}),
\label{eq:EPM}
\end{eqnarray}
where $a_{X,t_{i}}$ is a function that can be calibrated empirically and $v_{ph}$ the photospheric velocity derived from the Fe II $\lambda$5169 line absorption minima. Finally, the SN distance modulus is obtained using,
\begin{eqnarray}
\mu_{X,t_{i}}=m^{corr}_{X,t_{i}}-a_{X,t_{i}}+ 5log_{10}(v_{ph}),
\label{eq:EPM}
\end{eqnarray}
where $m^{corr}_{X,t_{i}}$ is the apparent magnitude corrected for Galactic, host galaxy extinctions and K-correction. Unlike the SCM method for which a color term is added in the SN standardisation, \cite{Rodriguez:2014} measured the host galaxy extinction using the colours and assuming that all SNe II have the same intrinsic colour. However, this assumption has recently been challenged by \cite{deJaeger:2018rzj} who demonstrated that intrinsic SN II colors are most probably dominated progenitor properties like its radius.

To calibrate the PMM zero-point (encapsulated in $a_{X,t_{i}}$), \cite{Rodriguez:2019} used four SNe~II with known distances from TRGB (SN 03hn, 05cs, 12aw, and 13ej). With only 9 SNe II with $cz$ $>$ 2000 km s$^{-1}$, they were able to derive $H_0$ value between 67.1 and 74.9 km s$^{-1}$ Mpc$^{-1}$. The large range of $H_0$ values is a source of concern for using the PMM as the $H_0$ values strongly depend on the filter chosen. For example, using the $B$ band, they obtained 74.9$^{+10.6}_{-9.2} \rm(stat)$ km s$^{-1}$ Mpc$^{-1}$ while using redder filter, this value decrease to 67.1$^{+5.4}_{-5.0} \rm(stat)$ km s$^{-1}$ Mpc$^{-1}$ in $H$ band (see Figure \ref{fig:HD_PMM}).\\

\noindent
\textbf{Pros}: Straightforward simple method. Smaller scatter in the Hubble diagram than the SCM.\\
\textbf{Cons}: Small sample size and affected by peculiar velocities. $H_0$ depends on the filters used. Host galaxy extinction estimated with colours. Needs to be calibrated. Never applied at high redshifts. \\
\textbf{Requirements}: Photometry (at least two bands) between 35--75 days since explosion, one optical spectrum during the plateau phase, a well-defined explosion date ($\sigma_{T_{exp}}<$5 days), calibrators (Cepheids, TRGB) in the same host.\\
\textbf{Future}: Increase the number of calibrators and SN in the Hubble flow.

\begin{figure*}[!t]
	\includegraphics[width=\textwidth]{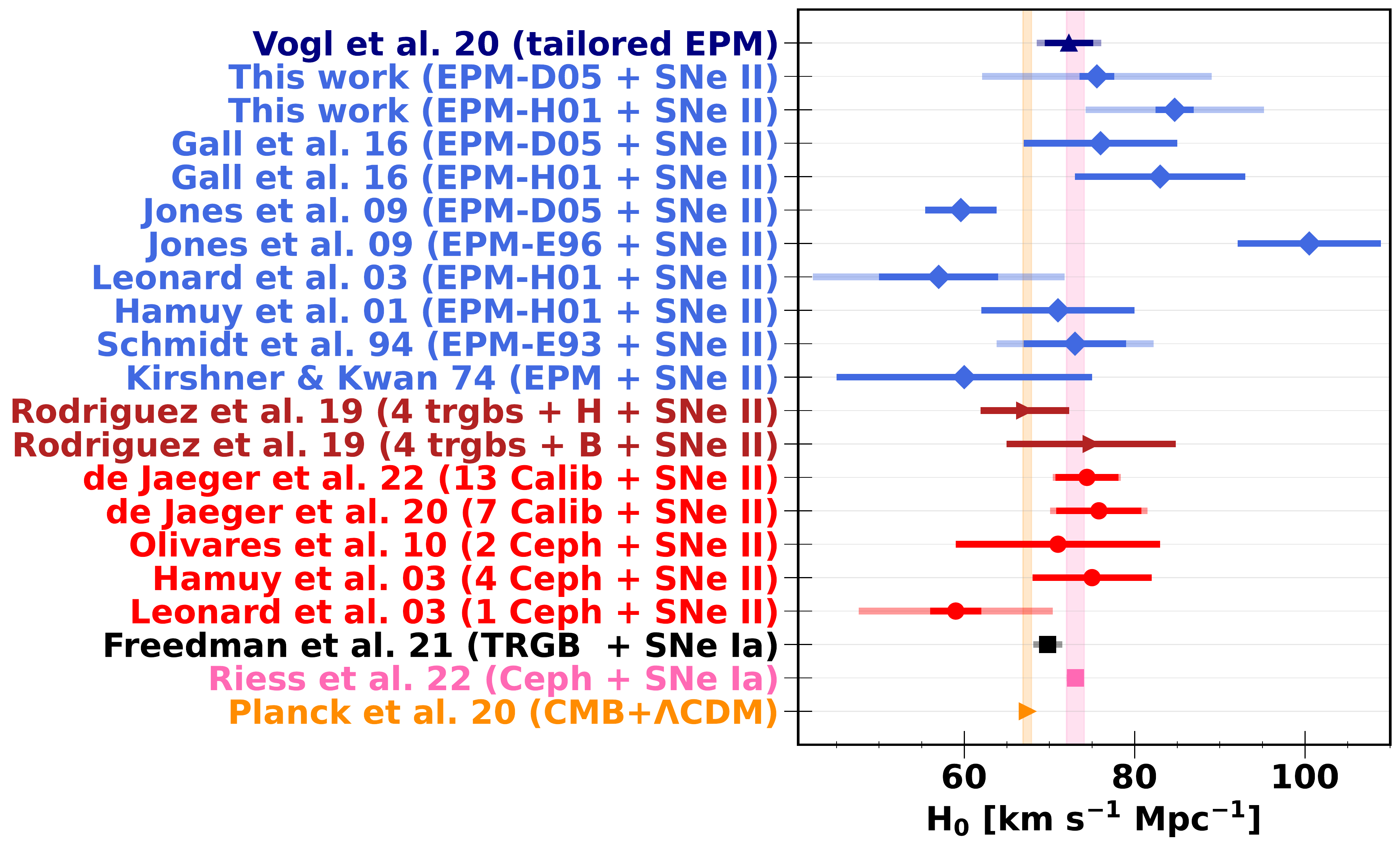}
 \caption{Comparison of $H_0$ derived using SNe~II, SNe~Ia calibrated with Cepheids (\cite{Riess:2021jrx}) and TRGB (\cite{Freedman:2021ahq}), and from the CMB anisotropies and lensing (\cite{Planck:2018vyg}). For the Type II supernovae measurements, we represent in red those using the SCM, in brown those from the PMM method, and in blue those using the EPM, including the tailored version. Light errors bars represent statistical and sytematic uncertainties in quadrature, while solid error bars are only statistical uncertainties.}
\label{fig:summary}
\end{figure*}

\section{Conclusions}

All $H_0$ measurements from SNe~II are summarized in Table \ref{tab:1} and presented in Figure \ref{fig:summary} in comparison to late-Universe Cepheid (\cite{Riess:2021jrx}) and TRGB  (\cite{Freedman:2021ahq}) estimates, and the early-Universe (\cite{Planck:2018vyg}) value from Planck. 
For the EPM values measured in this work and the tailored EPM by Vogl et al. (\cite{Vogl:2020thesis}) we considered the dispersion of all individual $H_0$ measurements as an estimate of the systematic uncertainty. 

Although not yet as mature as those using SNe Ia, different methods have been developed to standardize SNe~II to make them useful extragalactic distance indicators and, in turn, provide independent estimates of $H_0$.
Their main advantages over SNe~Ia are: (i) the extended knowledge on their progenitor systems and the physics behind their explosion mechanism; (ii) their larger volumetric rates; and (iii) their more direct connection with the local environment, which could potentially be used to improve their standardization. 
In addition, their constraints on $H_0$ are independent to those from SNe~Ia since they are affected by different systematic uncertainties.
Advantageously, some stand-alone methods exist that do not need calibrators as the distance ladder does.
The different methods summarized here, some of which do not depend on external calibrators, are more likely to be improved in the following years thus reducing the impact of systematic uncertainties in the determination of $H_0$.
Moreover, with more well-observed objects to be included that will reduce the statistical uncertainties, 
SNe~II will certainly 
contribute to understanding the reason behind the Hubble tension.



\begin{acknowledgement}
T.dJ acknowledges financial support from the French Centre National de la Recherche Scientifique (CNRS/IN2P3) and from the French Agence National de la Recherche project 21-CE31-0016.
L.G. acknowledges financial support from the Spanish Ministerio de Ciencia e Innovaci\'on (MCIN), the Agencia Estatal de Investigaci\'on (AEI) 10.13039/501100011033, and the European Social Fund (ESF) "Investing in your future" under the 2019 Ram\'on y Cajal program RYC2019-027683-I and the PID2020-115253GA-I00 HOSTFLOWS project, from Centro Superior de Investigaciones Cient\'ificas (CSIC) under the PIE project 20215AT016, and the program Unidad de Excelencia Mar\'ia de Maeztu CEX2020-001058-M.
\end{acknowledgement}



\end{document}